\input harvmac
\baselineskip=16pt

\def \tr {{\rm tr}}

\def \G {{\cal G}}
\def \m {\mu}

\def \ha{{\textstyle{1\over 2}}}
\def \four {{\textstyle{1\over 4}}}

\def \T {{\cal T}}
\def \tr {{\rm tr}}

\def \bet  {{\bar \eta}}
\def \et {\eta}
\def \H{{\cal H}} 

\def \te {\textstyle}

\def \D {{\Delta}}

\def \hS {\hat S}

\def \a {\alpha}

\def \four {{1\ov 4}}

\def \ep{\epsilon}

\def \cD {{\cal D}}

\def \x {\xi}

\def \k {\kappa} 

\def \g {\gamma}
\def \del {\partial}

\def \const {{\rm const}}

\def \b {\beta}

\def \p {\phi}
\def \m {\mu}

\def \vp {\varphi }
\def \t {\theta}
\def \Z {{\cal Z}} 
\def \td {\tilde }
\def \d {\delta}

\def \sm {$\s$-model }

\def \inv {^{-1}}
\def \ov {\over }
\def \four{{\textstyle{1\over 4}}}

\def \t {\theta}

\def \l {\lambda}

\def \lr { \lref}

 \def \h {\hat} \def \t {\theta} 
 \def \p {\psi} 
 \def \ch {\chi}
 \def \bc {\bar \chi}

\def \R {{\rm R}} \def \rf {\refs} 
\def \x {\xi } 
\def \G {{\cal G}}

\def \z {\zeta}
\def \T { { \cal T}}

\def \D {{\rm D}}
\def \sm {sigma model\ }

\def\const {{\rm const}} \def\m{\mu}\def\l
{\lambda}

\def \b {\beta}
\def \pp {\phi} \def \cc {{\rm c}} 
\lr \andre{
O.D.~Andreev and A.A.~Tseytlin,
``Two Loop Beta Function In The Open String Sigma
Model And Equivalence With String Effective Equations
Of Motion,''
Mod.\ Phys.\ Lett.\  {\bf A3}, 1349 (1988).}

\lr \an{O.D.~Andreev and A.A.~Tseytlin,
``Partition Function Representation For The Open
Superstring Effective Action: Cancellation Of Mobius
Infinities And Derivative Corrections To
Born-Infeld Lagrangian,''
Nucl.\ Phys.\  {\bf B311}, 205 (1988).}

\lr\ant{O.D.~Andreev and A.A.~Tseytlin,
``Generating Functional For Scattering Amplitudes And
Effective Action In The Open Superstring Theory,''
Phys.\ Lett.\  {\bf B207}, 157 (1988).}

\lr \NS { A.~Neveu and J.H.~Schwarz,
``Factorizable Dual Model Of Pions,''
Nucl.\ Phys.\  {\bf B31}, 86 (1971).}

\lr \mrt{R.R.~Metsaev, M.A.~Rakhmanov and
A.A.~Tseytlin,
``The Born-Infeld Action As The Effective Action In
The Open Superstring Theory,''
Phys.\ Lett.\  {\bf B193}, 207 (1987).}

\lr \gersha{ A.A.~Gerasimov and S.L.~Shatashvili,
``Stringy Higgs Mechanism and the Fate of Open
Strings,''
hep-th/0011009.}

\lr \gres {M.B.~Green and N.~Seiberg,
``Contact Interactions In Superstring Theory,''
Nucl.\ Phys.\  {\bf B299}, 559 (1988).
M.~Dine and N.~Seiberg,
``Microscopic Knowledge From Macroscopic Physics In
String Theory,''
Nucl.\ Phys.\  {\bf B301}, 357 (1988).
}

\lr\nss{ J. Scherk, ``Zero slope limit of the dual
resonance model", 
Nucl.\ Phys.\  {\bf B31}, 222
(1971).
A.~Neveu and J.~Scherk,
``Connection Between Yang-Mills Fields And Dual
Models,''
Nucl.\ Phys.\  {\bf B36}, 155 (1972).
J.~Scherk and J.H.~Schwarz,
``Dual Models For Nonhadrons,''
Nucl.\ Phys.\  {\bf B81}, 118 (1974).
T.~Yoneya,
``Connection Of Dual Models To Electrodynamics And Gravidynamics,''
Prog.\ Theor.\ Phys.\  {\bf 51}, 1907 (1974).
}

\lr \TTT { A.A.~Tseytlin,
``Ambiguity In The Effective Action In String
Theories,''
Phys.\ Lett.\  {\bf B176}, 92 (1986).
}

\lr\halp{K. Bardakci, ``Dual Models and Spontaneous Symmetry
Breaking", Nucl. Phys.
{\bf B68}, 331 (1974).
K. Bardakci and M.B. Halpern, 
``Explicit Spontaneous Breakdown in a
Dual Model", Phys. Rev. {\bf D10}, 4230  (1974);
 ``Explicit Spontaneous Breakdown in a
Dual
Model II: N Point Functions", Nucl. Phys. {\bf B96},  285 (1975).
K.Bardakci, ``Spontaneous Symmetry Breakdown in the Standard Dual
String
Model", Nucl. Phys. {\bf B133},  297 (1978).
}

\lr \tsey { A.A.~Tseytlin,
``On the tachyonic terms in the string effective
action,''
Phys.\ Lett.\  {\bf B264}, 311 (1991).
 }

\lr\dasg{S.~Dasgupta and T.~Dasgupta,
``Renormalization group analysis of tachyon condensation,''
hep-th/0010247.
}

\lr\mep{A.A.~Tseytlin,
``Partition Function Of String Sigma Model On A Compact Two Space,''
Phys.\ Lett.\  {\bf B223}, 165 (1989).
}
\lr\ame{
O.D.~Andreev, R.R.~Metsaev and A.A.~Tseytlin,
``Covariant calculation of the partition function of the two-dimensional 
sigma model on compact 2-surfaces,''
Sov.\ J.\ Nucl.\ Phys.\  {\bf 51}, 359 (1990).
}

\lr\mec{A.A.~Tseytlin,
``Mobius Infinity Subtraction And Effective Action In Sigma Model
Approach To Closed String Theory,''
Phys.\ Lett.\  {\bf B208}, 221 (1988).
``Conditions Of Weyl Invariance Of Two-Dimensional Sigma Model From
Equations Of Stationarity Of 'Central Charge' Action,''
Phys.\ Lett.\  {\bf B194}, 63 (1987).
}

\lr \TT { A.A.~Tseytlin,
``Vector Field Effective Action In The Open
Superstring Theory,''
Nucl.\ Phys.\  {\bf B276}, 391 (1986).
}
\lr \gw {D.J.~Gross and E.~Witten,
``Superstring Modifications Of Einstein's
Equations,''
Nucl.\ Phys.\  {\bf B277}, 1 (1986).}

\lr \ts { A.A.~Tseytlin,
``Renormalization Of Mobius Infinities And Partition
Function Representation For String Theory Effective
Action,''
Phys.\ Lett.\  {\bf B202}, 81 (1988).
 }
 
\lr \dor {H.~Dorn and H.J.~Otto,
``Open Bosonic Strings In General Background Fields,''
Z.\ Phys.\  {\bf C32}, 599 (1986).
}

\lr \wi { E. Witten, ``On background-independent
open-string field theory,'' Phys. Rev. {\bf D46}, 5467
(1992), hep-th/9208027.
``Some computations in background-independent
off-shell string theory,'' Phys. Rev. {\bf D47}, 3405 
(1993),  hep-th/9210065.
}

\lr \oku{ K. Okuyama, ``Noncommutative Tachyon from Background
Independent Open String Field Theory,''
 hep-th/0010028.}

\lr \ttt {A.A.~Tseytlin,
``Born-Infeld action, supersymmetry and string
theory,''
hep-th/9908105.
}

\lr \alv {L.~Alvarez-Gaume,
``Supersymmetry And The Atiyah-Singer Index Theorem,''
Commun.\ Math.\ Phys.\  {\bf 90}, 161 (1983).
D.~Friedan and P.~Windey,
``Supersymmetric Derivation Of The Atiyah-Singer Index And The
Chiral Anomaly,''
Nucl.\ Phys.\  {\bf B235}, 395 (1984).
}

\lr\corn{
L. Cornalba, ``Tachyon condensation in large
magnetic
fields with background independent string field theory,''
hep-th/0010021.} 

\lr \revi{A.A.~Tseytlin,
``Sigma Model Approach To String Theory,''
Int.\ J.\ Mod.\ Phys.\  {\bf A4}, 1257 (1989).
``Renormalization Group And String Loops,''
Int.\ J.\ Mod.\ Phys.\  {\bf A5}, 589 (1990).
}

\lr \ed {E.~Witten,
``D-branes and K-theory,''
JHEP {\bf 9812}, 019 (1998), 
hep-th/9810188.
}

\lr\km{D. Kutasov, M. Mari\~no, and G. Moore, ``Some
exact results on tachyon condensation in string
field theory,'' hep-th/0009148.}

\lr \sen {A.~Sen,
``Non-BPS states and branes in string theory,''
hep-th/9904207.
}
\lr \poll{ A.M. Polyakov, 
``Quantum geometry of bosonic strings,''
Phys.\ Lett.\  {\bf B103}, 207 (1981).
}

\lr \sha{A. Gerasimov and S. Shatashvili, ``On exact tachyon
potential
in open string field theory,'' JHEP {\bf 0010}, 034 (2000), 
hep-th/0009103.
}

\lr \kmm   {D.~Kutasov, M.~Mari\~no and G.~Moore,
``Remarks on tachyon condensation in superstring field
theory,''
hep-th/0010108.
 } 

\lr \sa {S. Shatashvili, ``Comment on the background independent
open string theory,'' Phys.\ Lett.\  {\bf B311}, 83
(1993),
hep-th/9303143;
``On the problems with background independence in string theory,''
hep-th/9311177.}

\lr\ftbi {E.S.~Fradkin and A.A.~Tseytlin,
``Nonlinear Electrodynamics From Quantized Strings,''
Phys.\ Lett.\  {\bf B163}, 123 (1985).
}

\lr\ft { E.S.~Fradkin and A.A.~Tseytlin,
``Quantum String Theory Effective Action,''
Nucl.\ Phys.\  {\bf B261} (1985) 1.
``Effective Field Theory From Quantized Strings,''
Phys.\ Lett.\  {\bf B158}, 316 (1985).
}

\lr\witli {K. Li and E. Witten,
``Role of short distance behavior in off-shell open-string
field theory,'' 
Phys. Rev. {\bf D48}, 853 (1993), hep-th/9303067.}

\lr\and {O.~Andreev,
``Some computations of partition functions and
tachyon potentials in  background independent
off-shell string theory,''
hep-th/0010218.
}

\lr\abo{A. Abouelsaood, C.G.~Callan, C.R.~Nappi and
S.A.~Yost,
``Open Strings In Background Gauge Fields,''
Nucl.\ Phys.\  {\bf B280}, 599 (1987).
}

\lr \hkm{ J.A. Harvey, D. Kutasov and E. Martinec, 
``On the relevance of tachyons,'' hep-th/0003101.}

\lr\mina{J.A.~Minahan and B.~Zwiebach,
``Field theory models for tachyon and gauge field
string dynamics,''
JHEP {\bf 0009}, 029 (2000), hep-th/0008231;
``Effective tachyon dynamics in superstring theory,''
hep-th/0009246.}

\lr \ishi{ S.~Samuel,
``Color Zitterbewegung,''
Nucl.\ Phys.\  {\bf B149}, 517 (1979).
J.~Ishida and A.~Hosoya,
``Path Integral For A Color Spin And Path Ordered Phase
Factor,''
Prog.\ Theor.\ Phys.\  {\bf 62}, 544 (1979).
A.~Barducci, R.~Casalbuoni and L.~Lusanna,
``Anticommuting Variables, Internal Degrees Of Freedom,
And The Wilson Loop,''
Nucl.\ Phys.\  {\bf B180}, 141 (1981).
J.L.~Gervais and
A.~Neveu,
``The Slope Of The Leading Regge Trajectory In Quantum
Chromodynamics,''
Nucl.\ Phys.\  {\bf B163}, 189 (1980).
H.~Dorn,
``Renormalization Of Path Ordered Phase Factors And Related Hadron
Operators In Gauge Field Theories,''
Fortsch.\ Phys.\  {\bf 34}, 11 (1986).
}

\lr \lud {
I. Affleck and A.W. Ludwig,``Universal noninteger
'ground state degeneracy' in critical quantum systems,''
Phys.\ Rev.\ Lett.\  {\bf 67}, 161 (1991).
} 

\lr \zam {A.B.~Zamolodchikov,
``'Irreversibility' Of The Flux Of The Renormalization Group
In A 2-D Field Theory,''
JETP Lett.\  {\bf 43}, 730 (1986).
}

\lr\hsue{C.S.~Hsue, B.~Sakita and M.A.~Virasoro,
``Formulation of dual theory in terms of functional
integrations,''
Phys.\ Rev.\  {\bf D2}, 2857 (1970).
M.~Ademollo, A. D'Adda, R. D'Auria, E.
Napolitano,
S. Sciuto, P. Di
Vecchia,
F. Gliozzi, R. Musto, F.
Nicodemi,
``Theory Of An Interacting String And Dual Resonance
Model,''
Nuovo Cim.\  {\bf 21A}, 77 (1974).
}

\lr\berg{
E.~Bergshoeff, E. Sezgin, C.N.~Pope and P.K.~Townsend,
``The Born-Infeld Action From Conformal Invariance Of
The Open Superstring,''
Phys.\ Lett.\  {\bf 188B}, 70 (1987).}

\lr\fri{
B.E.~Fridling and A.~Jevicki,
``Nonlinear Sigma Models As S Matrix Generating
Functionals Of Strings,''
Phys.\ Lett.\  {\bf B174}, 75 (1986).}

\lr \lov{C.~Lovelace,
``Stability Of String Vacua. 1. A New Picture Of The
Renormalization Group,''
Nucl.\ Phys.\  {\bf B273}, 413 (1986).
}


\lr\nonp{S.R.~Das and B.~Sathiapalan,
``String Propagation In A Tachyon Background,''
Phys.\ Rev.\ Lett.\  {\bf 56}, 2664 (1986);
``New Infinities In Two-Dimensional Nonlinear Sigma
Models And
Consistent String Propagation,''
Phys.\ Rev.\ Lett.\  {\bf 57}, 1511 (1986).
R.~Akhoury and Y.~Okada,
``Unitarity Constraints For String Propagation In The
Presence
Of Background Fields,''
Phys.\ Lett.\  {\bf B183}, 65 (1987).
C.~Itoi and Y.~Watabiki,
``Nonperturbative Effect In The Beta Functions And The
Equations Of Motion For String,''
Phys.\ Lett.\  {\bf B198}, 486 (1987).
Y.~Watabiki,
``Weyl Anomaly Of The Product Of String Vertex
Functions,''
Z.\ Phys.\  {\bf C38}, 411 (1988).
H.~Ooguri and N.~Sakai,
``String Loop Corrections From Fusion Of Handles And
Vertex Operators,''
Phys.\ Lett.\  {\bf B197}, 109 (1987).
R.~Brustein, D.~Nemeschansky and S.~Yankielowicz,
``Beta Functions And S Matrix In String Theory,''
Nucl.\ Phys.\  {\bf B301}, 224 (1988).
I.~Klebanov and L.~Susskind,
``Renormalization Group And String Amplitudes,''
Phys.\ Lett.\  {\bf B200}, 446 (1988).
S.~Elitzur, A.~Forge and E.~Rabinovici,
``Some global aspects of string compactifications,''
Nucl.\ Phys.\  {\bf B359}, 581 (1991).
A.R.~Cooper, L.~Susskind and L.~Thorlacius,
Nucl.\ Phys.\  {\bf B363}, 132 (1991).
J.~Schnittger and U.~Ellwanger,
``Nonperturbative conditions for local Weyl invariance on a
curved world sheet,''
Theor.\ Math.\ Phys.\  {\bf 95}, 643 (1993),
hep-th/9211139.
V.A.~Kostelecky, M.~Perry and R.~Potting,
``Off-shell structure of the string sigma model,''
Phys.\ Rev.\ Lett.\  {\bf 84}, 4541 (2000),
hep-th/9912243.
}

\lr \bank{
T.~Banks,
``The Tachyon potential in string theory,''
Nucl.\ Phys.\  {\bf B361}, 166 (1991).
}
\lr\sash{
A.M.~Polyakov,
``Directions In String Theory,''
Phys.\ Scripta {\bf T15} (1987) 191. }

\lr\call{C.G.~Callan, E.J.~Martinec, M.J.~Perry and D.~Friedan,
``Strings In Background Fields,''
Nucl.\ Phys.\  {\bf B262}, 593 (1985).
C.G.~Callan and Z.~Gan,
``Vertex Operators In Background Fields,''
Nucl.\ Phys.\  {\bf B272}, 647 (1986).
}

\lr\hon{
J. Honerkamp and G. Ecker, ``Application of invariant renormalization to
the non-linear chiral invariant pion lagrangian in the one-loop
approximation", Nucl. Phys. {\bf B35}, 481 (1971).
  J. Honerkamp, ``Chiral multi-loops", 
      Nucl. Phys. {\bf B36}, 130 (1972).       
D.H.~Friedan,
``Nonlinear Models In Two + Epsilon Dimensions,''
Annals Phys.\  {\bf 163}, 318 (1985).
}

\lr \kle { I.R.~Klebanov and A.A.~Tseytlin,
``D-branes and dual gauge theories in type 0 strings,''
Nucl.\ Phys.\  {\bf B546}, 155 (1999), 
hep-th/9811035.
}


\baselineskip8pt
\Title{\vbox
{\baselineskip 6pt{\hbox{
OHSTPY-HEP-T-00-025  
}}{\hbox
{   }}{\hbox{
}} {\hbox{
   }}} }
{\vbox{\centerline { Sigma model approach  }
\medskip
 \centerline {to  
 string theory effective actions with tachyons }
 }}
\vskip -20 true pt

\medskip
\centerline{   A.A. Tseytlin\footnote{$^{\star}$}{\baselineskip8pt
e-mail address:
tseytlin@mps.ohio-state.edu}\footnote{$^{\dagger}$}{\baselineskip8pt
Also at   Blackett Laboratory, Imperial 
College, London and  Lebedev  Physics
Institute, Moscow.} }

\smallskip\smallskip
\centerline {\it  Department of Physics }
\smallskip
\centerline {\it  The Ohio State University}\smallskip
\centerline {\it  Columbus, OH 43210-1106, USA}

\bigskip\bigskip
\centerline {\bf Abstract}
\medskip
\baselineskip14pt
\noindent
\medskip

Motivated by recent discussions of actions 
for tachyon   and vector fields related to tachyon
condensation in open string theory
we review and clarify  some aspects of their derivation 
within  sigma model approach. In particular, we
demonstrate that the renormalized partition function $Z(T,A)$
of boundary sigma model gives the effective action for massless
vectors
which is  consistent with string S-matrix and beta function, 
resolving an old problem with this suggestion in  bosonic string
case
at the level of the leading $F^2 (dF)^2$  derivative  
corrections to Born-Infeld action. 
We give manifestly gauge invariant definition of $Z(T,A)$
in non-abelian NSR open string theory 
and  check that 
its derivative reproduces 
the tachyon beta function in a particular scheme.
We also discuss derivation of similar actions for tachyon  and 
massless modes in closed bosonic and NSR (type 0) string theories.

\Date {November  2000}

\noblackbox
\baselineskip 16pt plus 2pt minus 2pt

\newsec{Introduction}

To try to 
address the question of vacuum structure of 
string theory 
it is natural to look for  a kind of field theory
 action which would interpolate 
between possible ground states, e.g., unstable  perturbative   
 and some stable  non-perturbative  one.
The original   S-matrix based method \nss\ 
of reconstructing string  effective 
action  order by order 
in powers of fields from on-shell  scattering amplitudes 
does not in principle allow one to find such an action.

It was suggested in \ft\  that  a useful 
framework   for an off-shell approach  
should be a generalized 2-d
\sm  partition function representing a 
generating functional \hsue\ 
 for correlators of string 
vertex operators given  by the Polyakov
   path integral \poll.
 The condensates of string fields are then 
 \sm couplings, and one may  hope to determine
the exact structure of the action without expanding 
in powers of them (e.g., expanding instead
 in derivatives of the fields).
 One advantage of this \sm approach is that 
 off-shell gauge  symmetries of low-energy expansions become 
 manifest.
 
 The precise definition of the 
 effective action (see \revi\ for a review)
 should be
 consistent with the string S-matrix near the 
 perturbative vacuum  and should also reproduce the 
 conditions of Weyl invariance of the \sm as its
  equations of  motion
   \rf{\lov,\call,\zam,\sash,\TT}.
In critical string theory,
    where, by definition, 
   one does not integrate over the conformal factor of the
   2-d metric, the form of the off-shell action 
   depends on a  Weyl symmetry gauge, but that dependence
   should disappear at the stationary points described by 2-d 
   conformal theories.

While this \sm partition function  approach  was successful 
for the massless string modes leading to covariant 
expressions to all orders in powers of gravitons and dilatons
in the closed string case and the  vector field strength 
in the open string case \ftbi, 
it produced   unfamiliar expressions  when applied 
to the tachyon field $T$. As was observed already in \ft, 
the expression for the partition function $Z[T]$ 
computed by expanding in derivatives of $T$
 has the following structure  in the critical bosonic string theory 
 (both in the closed string case on 2-sphere and open string
 case on the disc):
$Z= a_0 \int d^D x \ e^{- T }
  [ 1 + a_1 \a' \del^2 T + O(\a'^2)] $.
[For the closed string case
 this expression is given in 
 eqs. (39),(40) in \ft. In the critical open  
 string theory case 
 one is to omit an  additional  integral over the length 
 of the boundary  in the expression  following  eq. (54) in \ft.]
  The constant  $a_1$ was  renormalization scheme dependent 
  (logarithmically divergent before subtraction).
  Introducing $\Phi = e^{-T/2}$  and  properly tuning 
  $a_1$ one was able to reproduce the standard 
   tachyon kinetic term. The  meaning of that procedure 
   was, however, unclear.
   
Indeed, to  be able to make connection with
  the standard tachyonic amplitudes 
   it was  obvious that one should 
     expand in powers of $T$ 
   and not  derivatives of $T$  as the tachyon momentum
   should  be close to its mass shell value.
   The corresponding tachyon beta-function 
   then receives contributions which are 
    non-perturbative  in $\a'$ 
   \refs{\lov,\nonp,\tsey} and  which  are in agreement with the  
 tachyonic terms in the effective action reconstructed  
directly  from string amplitudes. The  form of the 
 ``tachyon potential", i.e. the   zero-momentum part
 of such action is inherently ambiguous  
 \rf{\bank,\tsey}, as one can always ``dress" any factor of
 $T$ by $\del^2$
 without changing  on-shell amplitudes.
 Thus one  needs some extra principle, not apparent at the level 
 of string S-matrix, to fix  this ambiguity.
 
 One could still  hope  that such  extra input
  was, in fact, 
 contained in the world-sheet \sm approach.
  This   was   effectively  vindicated  
  by the  recent derivations of  the tachyon potential
  in the open string theory (which were motivated by the study 
  of tachyon condensation on non-BPS 
   D-branes \rf{\sen,\hkm,\mina,\dasg}; for some
    early  studies of  tachyon condensation see \halp):
  $ e^{-T} (1 + T)$ in the bosonic string case
  \rf{\sha,\km} and $e^{-T^2}$ in the  
  NSR string case \rf{\kmm}.

  While the discussions in  
  \rf{\sha,\km,\kmm} were  presented 
 in the   framework of 
 Witten's   background-independent 
  open string field theory \rf{\wi,\sa}, 
  their  results  can be  
  obtained   directly  in the context of the  
 \sm approach as we shall review  below.
 
 The  idea is to return back 
 to the original boundary  \sm  \ft\ containing only the 
 tachyon and massless vector couplings.
 This model is renormalizable within the standard 
 derivative ($\a'$) expansion, i.e.  its set of couplings 
 is closed under perturbative RG  flow. 
 While one will certainly need to resum the  $\a'$ expansion 
 to be able to reproduce correct  interaction terms at 
  the standard tachyon vacuum point  $T=0$, 
 the low-energy expansion  (approximate in $\del T$ but 
  exact  in $T$)  
 may  be useful in order to  reveal the existence of a 
 new stationary point invisible  in perturbation
  theory near  $T=0$.
The main point  is that   low-energy expansion  near $T=\const$,
not applicable near  the standard perturbative vacuum, 
may be applicable near  the new one.
The hope is that the resulting action will  interpolate 
between the perturbative  and the  non-trivial vacua. 

  The central role  in the sigma model approach is played 
by the string  sigma model  partition  function  which is directly 
related to the generating functional 
$< e^{ \vp^{(in)}_a \cdot V_a }>$    for string acattering amplitudes.
 As was  argued in \rf{\TT,\ts},   the tree level 
 effective 
 action  $S[A]$ for the  massless vector field 
  should be  given simply
 by the {\it renormalized} partition function of the boundary
 sigma model,  as originally conjectured in \ft. 
 Renormalization of logarithmic infinities corresponds
  to   subtraction of  massless poles 
 \rf{\TT,\lov,\fri}, while elimination of  power divergences 
 by    a shift of the bare tachyon coupling 
 accounts for  a contribution of the tachyon poles in the
 massless amplitudes. When consistently implemented, 
 this renormalization procedure resolves   
 (as we shall explain in  Section 2.2 below) an apparent 
 contradiction between  $S[A]=Z[A]$  ansatz 
   and string S-matrix  found 
 at the level of the $F^2 (\del F)^2$ terms in \an.  

In the presence of a non-zero  (renormalized) tachyon 
background  the  $S=Z$ prescription requires a modification
in the case of {\it bosonic} string theory. Indeed,  
$ Z'={\del Z\ov \del T} $ does not vanish 
at the standard vacuum point $T=0$, so 
one needs to make a  subtraction
of the  derivative term 
$S[T]= Z[T] - T\cdot Z'[0] + ...\ $.
A  consistent modification of $S=Z$  satisfying
$S[0]=0$ was suggested 
in the context of the Witten's approach \rf{\wi,\sa}:
$\h S[T]= Z[T] + \b^T\cdot Z'[T]$, 
where $\b^T$ is the tachyon $\b$-function.
This form of   subtraction term is a natural one 
since it   preserves the property of RG invariance of the action.
This definition then leads to  the expression 
$e^{-T} (1+T)$ for the open bosonic string 
 tachyon potential \sha.
We derive the corresponding low-energy effective action 
in  Section 2.1 and  also generalize it to the presence of 
a constant $F_{mn}$ background.

The  complication  of  power divergences 
and associated 
 shift of  the tachyon coupling  
   is absent in the 
 case of world-sheet supersymmetric NSR string.
In particular, 
 while 
 in the bosonic string the tachyon couples linearly 
  to the fields of the massless sector, 
 it decouples from  them  in the NSR case (interaction terms
 are quadratic in $T$).
 Here the $S[A]=Z[A]$  prescription is manifestly
  consistent
\an\  and, moreover, 
 should apply also  to  the case 
 of a non-vanishing tachyon background \kmm.
 We discuss the NSR case in detail in Section 3, 
reproducing some of the results of \kmm. We also  
give a manifestly gauge invariant definition 
of the  partition function  in the general 
non-abelian case and demonstrate that the 
second-derivative part of the action  $S[T]=Z[T]$  
taken in a special scheme 
has its variation over $T$ 
 proportional to  the linear 
perturbative terms in the tachyon $\b$-function.
As in the bosonic case, we generalize this action
to the presence of constant $F_{mn}$ background, 
when the potential term becomes 
$e^{-T^2} \sqrt{\det(I + F)}.$

 One  of the lessons of 
  application of the \sm approach to  open string 
  theory is  that the ``global" covariant objects  
 defined by the \sm path integral
 --  partition
 function  or effective action -- may contain more 
 information than  a set of $\b$-functions (or, more precisely,
 Weyl anomaly coefficients) computed in a local 
 coordinate patch in  field (\sm coupling) space. 
Indeed, the information on a metric \zam\  on the
coupling space is effectively  encoded in $Z$. The effective action 
then may have  additional  stationary points  not seen
from  the $\b$-functions computed in  a ``standard" 
coordinate patch. This may happen 
  if the  field space metric $\k$ 
becomes degenerate at these points
  when described in terms of  ``standard"  coordinates.
For example,  the  field  space  may 
have a non-trivial topology, and thus may 
 need to be represented  by  several coordinate patches.

Since the  sigma model partition function  plays the central role 
in the open string case, leading to  the correct expression for the 
  tachyon potential, it 
 is natural to expect  that the same  should be 
true also in   the {\it closed} string case.
In section  4 we  apply the \sm approach 
to  discuss  the tachyon dependence of 
the effective actions  in {\it closed}  
 bosonic and NSR string theories.
 In the closed string theory 
   the effective action  for the massless modes 
   is determined by \sm partition function on 2-sphere 
   in the following way \rf{\mec,\mep,\revi}:
   $S[\l] = - ({\del Z \ov \del \ln \ep})_{\ep=1} =
    \beta^i{ \del Z \ov \del \l^i} $, 
    so that $S= \int d^D x \sqrt G  e^{-2\pp} ( \beta^\pp - \four G^{mn}
     \beta^G_{mn} ) = 
     \int d^D x \sqrt G  e^{-2\pp} ( D-26 + ...).$
The extra 
   derivative over the logarithm of 2-d cutoff
   (compared to the original $S=Z$ conjecture of \ft) 
    accounts 
   for the subtraction of the volume of 
   the M\"obius group  which is lorathmically divergent 
   in the 2-sphere case \mec\ (in both the bosonic and the fermionic string theories).
 We shall suggest that,   like in the open string case, 
    in the presence of a tachyon background 
    this relation  should  again 
    be modified 
   by subtracting  a  term  proportional to 
   $ {\del S \ov \del T}$  to satisfy the condition $\hat S[0]=0$. 
 The resulting tachyon potential
  is then  $- T^2 e^{-T}$. 

   No such modification is   necessary in 
   the closed  fermionic NSR (or type 0) string case, where we 
   argue that  (the NS-NS part of) the   effective action 
   depends on the tachyon field only through $(\del T)^2 $ and 
   $\del^2 T$, i.e. there  appears to be 
    { no} tachyon potential.

\newsec{Open bosonic string
}

Let us first  take a formal approach, 
 forgetting   about possible  connection to on-shell
string  S-matrix and  consider 
the partition function  for the  (Euclidean) 
boundary \sm  with two couplings 
$I=\int d \vp [{ 1 \ov  \ep}  T(x) + i A_m (x) \dot x^m  ]$.
 This theory is power counting renormalizable if one expands
 in powers of derivatives of $T$ and $A_m$, i.e. is closed under
 RG with all higher-derivative
  non-renormalizable  interactions (massive string modes)
  consistently decoupled.
 One  can then  ask  which is the functional 
 $S[T,A]$  (the boundary  analog \lud\ of 
the  c-function \zam)
 that reproduces the corresponding perturbative 
  $\b$-functions
 in the sense of $\del S\ov \del \l^i$ = $\k_{ij}(\l)  \b^j$, $ \ 
 \l^i=(T,A_m)$.
If we   decouple the tachyon
(solve for it in terms of $A_m$) 
the result should then be 
 the effective action consistent  with the 
$S$-matrix for the massless vector mode.
More generally, 
 such  ``effective action" functional  $S[T,A]$ 
may  represent  a natural off-shell extension, 
capturing  non-trivial 
behavior of string theory  far away from standard tachyonic
 mass shell.
Remarkably, this is indeed  what happens to be true, 
as indicated by the discussions in \rf{\sha,\km}.


It is useful to start  by 
 recalling  the  expression for the partition function
 (or  the generating 
 functional for tachyon and vector amplitudes 
 in open string theory on the disc) in the
 general 
  non-abelian case  \rf{\ft,\ftbi}
\eqn\zz{
  Z[T,A,\ep] =
   <\tr P \exp  \bigg(-  \int d \vp [  
   {\ep}^{-1}  T(x)  +  
   i A_m (x) \dot x^m  ]\bigg) > \ ,  
  }
 where  the averaging is done with the free string action in the
 bulk of the 
 disc  and $\vp \in (0, 2\pi)$ parametrizes its  boundary.
 Here $T$ and $A_m$ are Hermitian matrices in the 
 Chan-Paton algebra of $U(N)$.
We  consider the oriented string
 case relevant in  D-brane context and  define  the action 
 so that continued to the Minkowski signature it becomes real.
 $\ep={a\ov r} \to 0 $  is a dimensionless UV 
  cutoff, i.e. the ratio
 of  the short-distance 
  cutoff  and the radius of the disc.
 $<...>$ depends on $\ep$ through the propagator (see eq. (2.5) 
  below).
One  can make \zz\   more explicit 
  by using the well-known representation \ishi\ of 
path ordered exponent 
 in terms of the path integral over 1-d  anticommuting 
 fields $\et^a, \bet_{ b}$  
 in the fundamental  and antifundamental representations 
 of $U(N)$
 \eqn\zze{
  Z[T,A,\ep] =  < \int [d \et][ d\bet]
  \    \exp \bigg( -  \int d \vp [\bet_a\dot \et^a + 
   \bet_a ({ \ep}^{-1}   T^a_{\ b} (x)
    + i A^a_{\ bm} (x) \dot x^m)  \et^b ]\bigg)> \ .
  }
  The measure of integration  
     is assumed to contain the factor $ \bar \eta_c(0)
     \eta^c(2\pi)$.
 In the abelian case \zz\ is simply 
 \eqn\zz{
  Z[T,A,\ep] =
   < e^{ -  \int d \vp [  { \ep}^{-1} T(x)  +  
    i A_m (x) \dot x^m  ] }> \ .  
  }
The standard procedure  \rf{\ft,\ftbi,\ts}  to compute $Z$ is to first isolate the 
constant (``zero mode")  part of $x^m$ and integrate over the
internal points of the 
disc getting an  effective  1-d path integral for the 
boundary theory
 \eqn\qre{
 Z= a_0  \int d^D x \   e^{-W}\ , \ \ \ \ \
\ \  e^{-W} = < e^{-I } > = \int [d\x]\  e^{- { 1 \ov 4 \pi \a'}  \int
 \x G^{-1} \x \  -I } \ , }
  \eqn\jij{
  G(\vp_1,\vp_2) = {1\ov \pi}\sum^\infty_{n=1} {e^{-n\ep}
   \ov n} \cos n \vp_{12}   \ , \ \ \ \ \ \ \ \ \ 
    \vp_{12}=\vp_1 -\vp_2 \ , 
   }
 $$ I = \int d \vp \bigg[  { \ep}^{-1} ( T + 
  \ha  \x^m \x^k   \del_m \del_k T
 + {\textstyle {1\ov 6}} \x^m \x^k  \x^l   \del_m \del_k \del_l
  T +...)  $$ 
 \eqn\uio{+ \  i ( \ha \x^k  F_{km} + 
  {\textstyle {1\ov 3}}\x^k \x^l \del_l   F_{km} + ...)\dot
  \xi^m \bigg]
   \ .  }
We have  shifted  $x(\vp)   \to x +  \x(\vp) ,
 \  \int^{2\pi}_0 d \vp\ \x(\vp)=0$
 (so that $W$ contains contributions of 1-PI graphs only).
 In what follows we shall often 
set the inverse string tension $2\pi \a'$ 
to one, but 
the dependence on $\a'$  is easy to restore on dimensional
grounds,   $\del^k T \to (\sqrt {2\pi \a'} \del)^k T$, \
  $\del^k F_{mn}\to (\sqrt {2\pi \a'}
  \del)^k (2 \pi \a' F_{mn})$. 
  
If one  ignores all  higher than second  powers  in $\x$, 
  i.e. assumes that $\del_m\del_n T$ and $F_{mn}$ are constant, 
  the resulting path integral becomes gaussian and can be
 computed  explicitly, 
 as was done for $T=0$ in  \ftbi\ (see also \ttt) 
  and  including $\del_m\del_n T$ in \refs{\witli,\oku,
  \corn,\and}.
  
 One may ``resum" the  perturbative expansion 
 by including $F\xi\dot \xi$ term into  the propagator \abo;
 regularizing the final expression one gets \rf{\ts,\an}
 \eqn\exa{
  G^{mn} (\vp_1,\vp_2|F) 
 =\  { 1 \ov \pi} \sum^\infty_{n=1} {e^{-\ep n}\ov n}
  \big[\G^{mn}(F) 
\cos n \vp_{12} - i \H^{mn} (F)\sin n \vp_{12} \big] \ , }
\eqn\defi{ \G^{mn}(F) \equiv 
 [(I +   F)^{-1}]^{(mn)}= [(I -F^2)^{-1}]^{mn}
 = \d^{mn} + F^{mk}
F^{kn} + ... \ ,  }
$$
\H^{mn} \equiv - [(I +   F)^{-1}]^{[mn]}
=  [F(I -F^2)^{-1}]^{mn} = F^{mk} \G^{kn}    \ . $$
It is then straightforward
 to compute the leading terms in  expansion  of $Z$ 
in derivatives of $T$ and $F$ but to all orders in $F_{mn}$.

The model \zze,\zz\  is renormalizable in { derivative}
 ($\a'$) expansion, 
so that 
 $T$ and $A_m$ in \zze\ or \zz\  should be interpreted  as
  $\ep$-dependent  bare couplings which cancel all 
  the divergent terms, i.e.  \ts
  \eqn\re{
  Z[T(\ep),A(\ep),\ep] = Z_\R [T_\R,A_\R] \ ,  }
  where 
  \eqn\ta{
  T(\ep) = {\te{ 1 \ov 2\pi} }
  \ep \big[ 1 + h_1 (A_\R) \ln \ep + h_2(A_\R) \ln^2 \ep +
  ...\big] T_\R
    + \big[ k_1 (A_\R) + k_2(A_\R) \ln \ep + .... \big] \ , }
    \eqn\aa{
  A(\ep) = A_\R   +    f_1( A_\R) \ln \ep +  f_2(A_\R) \ln^2 \ep + ...\ .
  }
 Here the renormalized fields
 are defined at point $2\pi r$ 
 and $h_i$ contain differential operators acting on $T_\R$. 
 For example, it is easy to show that in the abelian case 
 and  for constant  
  $F_{mn}$  background 
 \eqn\eas{
  h_1 (A)T = 
  {\te { 1 \ov 2\pi} } \G^{mn}( F)  \del_m \del_n T \ , \ \ \ \ 
    \  [f_1( A)]_k = {\te{ 1 \ov 2\pi}}   \G^{mn}(F)  \del_m F_{nk}\ . 
     }
$f_1$ represents  the 
 ``Born-Infeld"  $\beta$-function \rf{\abo,\dor}.
 The 
 inhomogeneous term in  \ta\  \ts 
 \eqn\tem{
   k_1 = -{ \te {1 \ov 4 \pi} }\ln  \det ( \d_{mn} +  F_{mn} ) 
   =- { \te { 1 \ov 8 \pi} }   F^2_{mn} +  O(F^4)  \ , }
  corresponds to a shift  of the  bare 
tachyon needed to be done to 
absorb the $F_{mn}$-dependent 
 linear divergence  
    appearing in the 
 computation of  $Z$ for $F_{mn}=\const$
 leading to   the BI action \rf{\ftbi,\ts} 
\eqn\coo{
W= b_0\ln  \det ( \d_{mn} +  F_{mn} ) \ , \ 
\ \ \ \ \   \ \   
b_0= \sum^\infty_{n=1}  e^{-2\ep n} = {\te {  1 \ov 2 \ep}} - 
 { {\te {1 \ov 2} }}+ O(\ep)\ . }
 If one subtracts the term \tem\ from the beginning, 
 it  will  not appear in the corresponding 
  $\b^T$-function.
  This is a scheme-dependent \TTT\   property, as
 one  can of course induce a similar  term back 
 by a field redefinition, $T \to T + f(F)$.
[Similar
 inhomogeneous term 
 does not appear in the tachyon $\b$-function in
 the closed string case  if one uses the natural scheme 
 in which the general covariance of the theory is manifest
  \tsey.]
  This  scheme is fixed by the requirement that the corresponding
  effective action 
  with $T=0$ and $F_{mn}=\const$ is given simply by the BI 
  action  (which itself is related, via D-brane action connection,
  to basic reparametrization  symmetry of the underlying string 
  theory).
The subtraction of   \tem\  to be done 
 in the bare partition function 
 will play an important role in Section 2.2  below.

The renormalized value of the partition function takes the form
(here and in what follows 
we omit subscripts ``R" on the renormalized value of 
 $Z$ and the fields)
$$
Z=a_0\int d^D x \ e^{ -  T} \ 
 \sqrt {\det (\d_{mn} +  F_{mn})}\bigg[
1 +  a_1 \a'  \G^{mn}(F)  \del_m \del_n T $$  \eqn\subs{ +  \ 
 {\cal F}^{kmn acd}(F) \del_k  F_{mn} \del_a  F_{cd} 
+ O (\del^4T,\del^4F^k) \bigg] 
\ ,  } 
where $
  {\cal F}^{kmn acd}(F) \sim F^2 + F^4 + ...$ \an\
  and  $ \a' = { 1 \ov 2 \pi}$ (i.e. $2\pi\a'=1$).

 The coefficient $a_1$ is  logarithmically divergent 
 before renormalization
 ($ \pi  G(\vp,\vp) =
  \sum^{\infty}_{n=1} { 1 \ov n}  e^{-\ep n}
  = - \ln \ep  + O(\ep)$)
  and thus is scheme dependent, 
 i.e. its value   can be changed by a field redefinition
 \rf{\TTT,\ts}.
 

 \subsec{Tachyon action }
 Let us first set $F_{mn}=0$ and consider the dependence of
 $Z$  \subs\ 
 on $T$
 \eqn\subse{
Z=a_0\int d^D x \ e^{ -  T} \ 
 (1 + a_1\a' \del^2 T + ...) =  
 a_0\int d^D x \ e^{ -  T} \ 
 [1 + a_1\a' ( \del  T)^2  + ...] \ ,  }
 i.e. 
 \eqn\hoot{
 Z=a_0\int d^D x\  [\ \Phi^2 + 4 a_1\a' (\del \Phi)^2  + ...]
 \ , \ \ \ \ \ \    \Phi \equiv e^{-T/2} \ . }
 This expression 
 which  looks like  an  action for a massive field
  with  $m^2=  (4 a_1\a')^{-1}$ was first found in \ft. 
  However, its derivative  does not
  vanish for $T=0$, i.e. does not  reproduce the 
    perturbative  tachyon coupling 
      $\beta$-function  which is  given, 
    to all orders in the  $\a'$ expansion, simply by 
(cf. \eas) 
 \eqn\bbb{ \beta^T = 
  -T - \a'  \G^{mn}( F)  \del_m \del_n T  = 
- T -  \a'  \del^2 T  + O(F\del\del T) \ . } 
This suggests that in the {\it bosonic} open 
 string theory,  
the  definition of the effective action 
as the 
renormalized \sm partition function $Z$ \rf{\ft,\ts}
 needs  a modification
when the tachyon background is non-zero.
The required 
 refinement of the $S=Z$  relation was suggested in  
 \rf{\wi,\sa}:  to define an  action
functional which will be stationary at conformal points
 one is to add   an  extra derivative term
\eqn\rer{
\hat S= S + \beta^T\cdot  {\delta  \ov \delta T} S
 =  Z + \beta^T\cdot  {\delta  \ov \delta T} Z
\ . }
The second  subtraction term is a natural one 
as it preserves the property of RG invariance of the action.
Note that  $\hat S$ and $Z$ are 
 equal at the stationary points of 
$\hat S$. Also, 
\eqn\rera{
\hat S[T]= S[T +  \beta^T] + O((\beta^T)^2)
\ = \ e^{\beta^T\cdot  {\delta  \ov \delta T} }\ Z[T] 
 +
O((\beta^T)^2) \ , }
i.e. changing from 
$S=Z$ to $\hat S$ may 
look like  a field redefinition  of $T$. This 
redefinition is,  however, 
singular in the case of the tachyon coupling.

In general, if   for a  set  of fields (\sm couplings)  $\l^i$ 
which is closed  under the RG  one has  \rf{\wi,\sa}
\eqn\rte{
 \hat S =  Z + \b^i \del_i Z \ , \ \ \ \ \ \ \ \ \ \ \ \  
\del_i\hat  S = \k_{ij}  \b^j \ , }
then
\eqn\rteq{
\del_i \hat S 
 = \del_i Z + \del_i \b^j \del_j Z + \b^j \del_i \del_j Z \ ,
} 
\eqn\rtes{
 (\k_{ij} -\del_i \del_j Z) \b^j
 = (\d^j_i + \del_i \b^j ) \del_j Z \ , }
 so that $\del_j Z=0$ may not imply $\b^j=0$ 
 if the ``shifted" 
  matrix  of anomalous dimensions 
  $\d^j_i + \del_i \b^j$  is  degenerate in some limit (e.g.
  at low momenta). 
 This is precisely  what happens in the  tachyon field  case
 (cf. \bbb), so  that  the modification 
 \rer\  is important here.

Using \subse,\bbb\  we find  that \rer\ is given,
to the
 leading order in derivatives
of  $T$, by  (cf. \rf{\sha,\km}) 
\eqn\lea{
\hS= a_0\int d^D x \ e^{ -  T} \ 
 \bigg[1 +  T +  (1-a_1)\a'  (\del T)^2 + a_1 \a' T (\del T)^2   +
 O(\del^4 T) \bigg] 
 \ . }
 Choosing   a  {\it special}  scheme where 
 \eqn\schi{
 a_1 = \ha \ , }
  one  finally gets 
 \eqn\leas{
\hS= a_0\int d^D x \ e^{ -  T} \ 
 \bigg[(1 + T) (1 +  \ha \a' \del^m T \del_m  T ) + O(\del^4 T) \bigg] 
 \ .  }
 Then 
 \eqn\rwt{
 {\delta \hS \ov \delta T}
 = \ a_0\ e^{ -  T} \ 
 [- T  - \a' \del^2 T  -  \a'T\del^2 T + \ha\a'  T (\del T)^2  +
 O(\a'^2\del^4 T) 
  ] \ , }
 which is indeed proportional 
 to the $\beta^T$-function  \bbb\  to the 
 leading order in $T$
 \rf{\sha,\km}.
 Here $e^{-T}$ should be interpreted as the field space metric 
 $\k_{TT} (T)$ in 
 \rte\  and the  non-linear terms in $T$ should be
  redefinable away (within $\a'$ or derivative expansion
  \bbb\ should be the exact expression 
  for the $\b^T$-function).
 Eq. \rwt\ has two obvious zeros: $T=0$ and $T=\infty$
 with the second one related to the tachyon condensation 
 \rf{\hkm,\km}.
Another 
  solution $T(x) = a + u x^2$  with finite constants 
  $a,u$ \km\ is  an artifact  of $\a'$ expansion
  (it does not correspond to a conformal 2-d theory).
[$T= u x^2_a$ with $u\to \infty$ which is a stationary point of $\hat S$ 
but  does not directly
solve $\beta^T=0$ in \bbb\   does  define  a CFT since 
it corresponds  to $\delta(x_a)$, i.e. Dirichlet boundary conditions in $x_a$
direction.]

 Including   $F_{mn}=$const  background  we  get
 the following generalization of  \subs,\lea
 \eqn\leak{
\hS= a_0\int d^D x \ e^{ -  T} \ \sqrt {\det (\d_{mn} +  F_{mn})}
 \bigg[1 +  T +  \ha \a' \G^{mn} (F)   \del_m \del_n T 
  +  ... \bigg] 
 \ , }
 or 
 $$ 
\hS= a_0\int d^D x \ e^{ -  T} \ \sqrt {\det (\d_{mn} + 2 \pi \a' 
  F_{mn})}
 \bigg[(1 +  T ) [1 + \  \ha \a' \G^{mn} (2 \pi \a'F)\   
  \del_m T \del_n T] $$  \eqn\leaks{
  +\   O(\a'^2\del^4 T,\a'^2\del^2 F) \bigg] 
 \ ,   }
 where we restored the full dependence on $\a'$.
 The variation of this  action 
 is proportional  the $\b^T$-function
 \bbb\ with
 $\del^2 T$ replaced by  $  \G^{mn}(F) \del_m \del_n T$ 
 (in agreement with \eas)  and which does not contain   
 the inhomogeneous term \tem.

One may raise the  question of 
 why the action \leaks\ is consistent with the string
S-matrix which  contains a  non-vanishing 
tachyon-vector-vector  amplitude. The latter   can be 
 reproduced   by the $T F^2_{mn}$ term 
 in the  effective action but such term 
  is not present 
 in \leaks. However, the term $-\a' F^2_{mn}\del^2 T$
 (or $-2\a' T \del_k F_{mn} \del_k F_{mn}$) 
 leads to the same on-shell 3-point amplitude since  for 
 the on shell tachyon  $\a'\del^2 T = -T$.
 Such higher derivative term is indeed present in 
 \leak\ or \leaks. 
 The corresponding $ \del_k F_{mn} \del_k F_{mn}$
 term in \rwt\ or  in the tachyon $\b$-function 
 is not visible in the $\a'$ expansion 
 but can be 
  reproduced if one expands in powers of the 
  fields instead of
 powers of derivatives and sums 
 all orders in  $\a'$ 
 (see \nonp).
Let us note that this case is
  completely analogous 
 to the $ R T$ vs.  $R^2_{mnkl} T$ contribution
 (giving the same on-shell graviton-graviton-tachyon amplitude)  
in the closed string  effective action   discussed 
in  \tsey.

\subsec{Vector field action}

Let us now  set  $T$ to zero 
and consider the dependence of $Z$ on 
the vector field $A_m$.
In view of the arguments given in \rf{\TT,\ts,\an}
 the effective action 
$S[A]$  which reproduces the string S-matrix 
and is also consistent  with the expression for the vector 
field  $\beta^A$-function 
in  the boundary  \sm  should be  given simply by the 
{\it renormalized}  value of the \sm partition function 
\eqn\sss{
S[A_\R] = Z_\R[A_\R, T_\R=0] \ . } 
We shall again omit subscripts ``$\R$" below.

This relation    passes a number of non-trivial 
tests. In the abelian case,  
for  $F_{mn}=\const$ 
one finds that $Z$ is equal to the BI action 
\ftbi\ whose derivative over $A_m$   is  indeed 
to be proportional to the 
leading one-loop term  $- \a' \G^{mn}(F) \del_m F_{nk}$ in the 
$\beta^A$-function \rf{\abo}.
The $F^4$ term in the expansion of the BI action is 
also in agreement with the string 4-point amplitude 
\rf{\TT,\gw}. In the non-abelian case, 
the direct computation  of $Z(A)$ defined by \zz\ 
gives, after a renormalization,  \ts\
\eqn\giv{
Z= a_0\int d^D x\  \tr \bigg[
1 - (2\pi \a' )^2  [ \four  F^2_{mn} + {\te{ 2 \ov 3} }\a' F_{mn} F_{nk} F_{km}
 +  d_1 \a'( D_m F_{mn})^2 ] + O(\a'^4)
 \bigg] \ ,  }
 where $d_1$ is scheme-dependent. This coincides  
  with the non-abelian  $F^2+F^3$ terms in 
 the action 
 reconstructed from the bosonic string S-matrix \rf{\nss,\TT}.

 There was, however, a  problem with  $S=Z$ ansatz \sss\   
 in the {\it bosonic}  string case
  pointed out in \an.
 The direct computation  of the leading derivative 
  $\del F$ dependent 
  terms in the abelian  partition function \zz\ 
   gave
  the following expression ($2 \pi \a'=1$) \an\ 
$$
Z= a_0\int d^D x  \bigg[1 + \four F^2_{mn}  - {\te { 1 \ov 8}}  
  [ (F_{mk}F_{kn})^2 - \four (F^2_{mn})^2 ] $$
\eqn\giv{  + \
   b_1 F_{kl}F_{kl} \del_a F_{mn} \del_a F_{mn} 
+ b_2  F_{kl}F_{lm} \del_a F_{mn} \del_a F_{nk}
 + b_3 F_{la}F_{lb} \del_a F_{mn} \del_b F_{mn}
+ O(  \del^2 F^6)\bigg] ,   }
where 
\eqn\coe{ b_1 =   { { 1 \ov 24 \pi}}\ , \ \ \ 
  b_2 = -  {{ 1 \ov 6 \pi}}\ , \ \ \ 
  b_3 =  { { 1 \ov 12 \pi}}\ , 
  }
  and  we have  ignored  all 
   terms  which have scheme-dependent coefficients
  (i.e. vanish on $\del_m F_{mn}=0$).
At the same time,  both the  string 4-point amplitude \an\
and the 2-loop \sm  $\b^A$-function calculation \andre\ 
  led
to the action \giv\ with the coefficients 
  \eqn\coeg{ b_1 =   - { { 1 \ov 48 \pi}}\ , \ \ \ 
  b_2 = -  {{ 1 \ov 6 \pi}}\ , \ \ \ 
  b_3 =  { { 1 \ov 12 \pi}}\ ,   
  }
 i.e.  with  the {\it same}  $b_2$ and $b_3$ 
  but with $b_1$ differing
  by factor $- \ha$.
  This apparent 
  disagreement of the $S=Z$ ansatz  with the 
  S-matrix and the 
  $\b^A$-function was attributed in \an\ to the  presence of the
  tachyon in the bosonic string.  
  The  tachyon poles are formally
  expanded in momenta in the derivation of the massless field 
  effective action from the string S-matrix,  
 and  it was suggested that 
   the corresponding  subtraction of 
  power divergences in $Z$  may be   hard to implement 
   unambiguously.

  This problem has, in fact, 
     a very  simple resolution  implied by
     the definition  \sss:  a  particular 
  ``tachyon-related" $F^2\del F\del F$  term  was missed 
    in  \an\  since  the tachyon field 
   there was set equal to zero {\it before}  properly 
   subtracting the  linearly divergent $F$-dependent 
    term 
    \tem. This subtraction effectively accounts for 
    the contribution  coming from the 
   expansion of the tachyonic pole in the 
   4-vector amplitude  which is included   in the effective
   action  if  it is  reconstructed from the 
   S-matrix.
   Once  this extra term  is added, 
  the agreement between  $S[A]$ and $Z[A]$ is indeed 
  restored!

 As was stressed  above, the effective action 
 should be given  by the {\it renormalized} value of $Z[A]$.
 The bare partition function has the same structure as \subs\
 (with $T\equiv T_\R  \to {2\pi \ov \ep} T$) 
 but in addition it  contains the linearly  divergent term in the
 exponent  $ e^{ {2\pi \ov \ep} k_1} $
   (see \coo,\tem). Shifting the bare tachyon 
 \ta\ to absorb this linear divergence (and then setting 
 $T_\R=0$)
  introduces  extra 
 $\del F$-dependent terms effectively originating  from the 
 $\del^2 T$ term in \subs.
 Explicitly,  applying  the {\it same} computational 
scheme  ($\ep$-regularization in \jij\ plus minimal subtraction)
that was  used in the original computation of
\giv,\coe\ in 
\an\ 
one finds the following expression
 for  $Z(T,A)$ in \qre,\jij\  
(in the form described below 
  this computation  was done by  O. Andreev; it corrects  
  the original version of this argument)
\eqn\yytr{ Z= \int d^D x\  e^{ -{ 2\pi \ov \ep} T} 
\big[1 + k_1(\ep) \del^2 T  + k_2(\ep)  F^2_{mn}
 + k_3 (\ep)  F^2_{mn} \del^2 T
 +k_4(\ep)  F^{km} F^{kn} \del_m \del_n T + O(\del^4)\big]\ ,}
where  $2\pi\a'=1$  and 
\eqn\chui{
k_1 = - \ep^{-1} I_1 (\ep)\ , \ \ 
\ 
k_2=  - \ha I_0 (2 \ep) \ ,\ \ 
 \ k_3=  \ha \ep^{-1} I_0(2\ep) I_1(\ep)\ ,\ \ 
\  k_4= \ep^{-1} I_1 (3\ep)\ , }
and 
\eqn\uyt{  I_0(\ep) \equiv 
\sum^{\infty}_{n=1}   e^{-\ep n}=
 (e^{\ep} - 1)^{-1} = {\te{ 1\ov  \ep }} - \ha + {\te {\ep \ov 12}} + 
  O(\ep^2)\ , \ \ \ } \eqn\uuyyt{
  I_1(\ep) \equiv 
\sum^{\infty}_{n=1}  { 1 \ov n}  e^{-\ep n}= - \ln 
 (1- e^{-\ep}) = - \ln \ep  + \ha \ep  +  O(\ep^2)\ . }
 To cancel the leading  power divergences 
 in the $F^2_{mn}$ and $F^2_{mn} \del^2 T$
 terms one is to  shift $T\to T - { 1 \ov 8 \pi} F^2_{mn} $
 in the exponent.  Using the {\it minimal}  subtraction 
 to absorb the singular logarithmic 
 parts of the coefficients $k_i$ 
 into the bare couplings  $T$ and $A_m$  one is left with 
 the  following values for their finite ($\ep\to 0$) 
 parts: 
 $(k_2)_{{\rm fin}} ={ 1 \ov 4} , \
 (k_3)_{{\rm fin}} =- { 1 \ov 8} , \ \
 (k_4)_{\rm fin}  =  { 3 \ov 2}$. The additional  
  $F^2 \del\del F^2$ terms coming from \yytr\ 
   that should be added to \giv\ are 
  then 
 \eqn\ytr{\Delta Z= a_0\int d^D x  \ 
  (  {\te { 1 \ov 16 \pi }}     ) 
   \big[   \four  F^2_{kl}  \del^2 (F^2_{mn}) - 
       3 F^2_{pq}   \del_m \del_n ( F_{mk} F_{nk}) \big]
   \ .} 
Using $\del_{[k } F_{mn]} =0$  and dropping 
terms proportional to the 
equations of motion so that  
 $\del_n  F_{mk} \del_m F_{nk} = 
\ha  \del_k F_{mn} \del_k F_{mn}   + O (\del_m F_{mn}), 
$
we find
\eqn\ytre{\Delta Z= a_0\int d^D x  \ 
  ( -   {\te { 1 \ov 16 \pi }} )  
  F_{pq} F_{pq} \del_k F_{mn} \del_k F_{mn}
   \ .} 
   Thus this  ``tachyonic"
   correction contributes only   to the  coefficient 
   of the first  $F^2 \del F\del  F$ term   in \giv\ 
   \eqn\cha{
b_1 \  \to \ b_1  -  {{ 1 \ov 16 \pi} } \  ,  }
  changing  it   precisely  into the  $b_1$  in \coeg.
  This 
implies   the {\it equivalence} of the leading $\del F$ 
derivative terms in 
the effective actions  obtained (i)  from 
the partition function,  (ii) from  the  string S-matrix, 
and (iii) from the $\beta^A$-function.
  
\newsec{Open   NSR string 
}
Ignoring first the tachyon, the  analog of the 
partition function \zz,\qre\  which is the generating functional 
for massless vector scattering  amplitudes is given by
\rf{\TT,\ant,\an}
$$
 Z(A) = < \tr\ P \exp \bigg(- i\int d \vp \ [ \dot x^m A_m (x) - \ha
 \psi^m\psi^n F_{mn}(x)]\bigg) >   $$
 \eqn\acti{ 
 = \int d^D x  < \tr\ P 
  \exp \bigg( -i \int d \vp \ [ \dot \x^m A_m (x + \x) 
- \ha \psi^m\psi^n F_{mn}(x + \xi)   ]\bigg) > \  , }
where the averaging is done with the free string propagator 
 restricted to the boundary of the disc, 
i.e.  with 
the effective 1-d boundary action 
  $I_0= { 1 \ov 4 \pi \a'} \int (\x G\inv \x + \psi K\inv 
\psi )$ 
with periodic $\xi^m(\vp)$ and  antiperiodic $\psi^m(\vp)$.
The bosonic Green function in \jij\ is now supplemented by 
the fermionic one
\eqn\tytr{
G(\vp_1,\vp_2) = {1\ov \pi}\sum^\infty_{n=1} {e^{-n\ep}
   \ov n} \cos n \vp_{12}   \ , \ \ \ \ \ \ \ \ 
 K(\vp_1,\vp_2) =   {1\ov \pi}\sum^\infty_{r=1/2 } {e^{-r\ep}
  } \sin r  \vp_{12}  \  .  }
As discussed in \rf{\an}, the $\ep\to + 0$ regularization 
preserves  underlying 
1-d  supersymmetry 
(which is spontaneously broken by the antiperiodic 
boundary condition on $\psi^m$, i.e. is an ``asymptotic"
symmetry).

 $P$ in \acti\ stands for 
the standard path ordering. The  contact  
 $[A_m,A_n]$ term in $F_{mn}$ 
implying manifest non-abelian gauge invariance 
of the resulting amplitudes 
can be derived  \refs{\ant,\an}
from   the  contact   terms
in the supersymmetric theta-functions  in the 
manifestly  1-d supersymmetric definition 
of the  path ordering (see also below).

To include the tachyon field, one may  start with the 
standard NS \NS\ 
vertex operator $\int d\vp\ \psi^m \del_m T(x)$.
This coupling cannot, however, be added directly into 
the exponent in \acti\  as $\psi$ is Grassmann  
while $T$ is not (integrating $\psi^m$ out would 
leave no dependence on $T$). To get a non-zero 
answer for the correlators one  is  to properly order 
the interaction vertices. 
 A simple way to do that,   as 
  suggested in \rf{\ed} and  elaborated on  in 
  \rf{\hkm,\kmm},  is to 
 introduce a  non-dynamical 1-d anticommuting, real, antiperiodic
   field 
 $\z(\vp)$,  and to  add   to the  action 
 the following terms 
 $\int d\vp [ \z \dot \z  + i \z \psi^m \del_m T(x)]$.

\subsec{General non-abelian case}

More precisely, to automatically include the  contact 
terms which will  make the non-abelian 
gauge invariance explicit, one is to insist 
on manifest world-sheet supersymmetry  of the \sm 
 interaction terms 
\rf{\gres,\ant,\an}. As in \rf{\alv,\an} here
 this  is accomplished   by 
 replacing $x^m$ by the 1-d scalar superfields 
 $\h x^m = x^m + \t \p^m$, and  the $U(N)$ ``quarks" 
 $\eta^a, \bet_a$ in \zze\  and  the new variable 
 $\z$ by  the ``spinor" superfields 
 $\h \et^a = \et^a + \t \ch^a, \ \  \h \bet_a =\bet_a +  \t \bc_a $ 
 and $\h \z = \z + \t f$. 
 This ensures  the 1-d supersymmetry of the path ordering 
 \rf{\ant,\an}.
 The resulting partition function 
 is given by the path integral over 
 $\h x^m, \h \et , \h \bet , \h \z$ similar to \zze\
 with the interaction part of the action now  being 
\eqn\repl{
 I=- \int d\vp d \t \bigg( \h \bet D \h \et +  \h \z D \h \z
 +
 i \h \bet
 \big[\ \h \z\ \T(\h x)   +  
 A_m (\h x) D \h x^m \big]   \h  \et\bigg)\ , 
 }
 where $D\equiv \t \del_\vp - \del_\t$
 and we suppressed  the $U(N)$ indices on
 $\h \et^a, \h \bet_a$ and the fields $\T^a_{\ b}, A^a_{\ b
 m}$.
 Here  $\T$ is the bare tachyon, i.e.
 $\T \sim { 1 \ov \sqrt \ep} T $.

The component form of \repl\ is 
 $$
 I=\int d\vp  \bigg[ \bet \dot \et  + \z \dot \z + f^2 + 
  \bc \ch  
 + i \bet (  \T \z  + A_m \psi^m) \chi
 - i\bc ( \T  \z  - A_m \psi^m) \eta   $$
 \eqn\rep{
  +\ i  \bet [ f \T - \z \psi^m \del_m \T
   +   (A_m \dot x^m -  \psi^m \psi^n \del_m A_n ) ]\eta \bigg]
\ .   }
This action  is manifestly 1-d supersymmetric, but 
its  non-abelian gauge invariance 
becomes apparent only after integrating over 
the auxiliary fields $\ch,\bc$
  \eqn\repo{
 I=\int d\vp  \bigg[ \bet \dot \et  + \z \dot \z + f^2  + 
 i \bet [   f \T  - \z \psi^m \D_m \T     +
(  A_m  \dot x^m - \ha \psi^m \psi^n F_{mn} )] \eta \bigg]
\ ,    }
where 
$$
 \D_m \T = \del_m \T + i [A_m, \T] \ , \ \ \  \ \ 
  F_{mn} = \del_m A_n - \del_n A_m + i[A_m, A_n] \ . $$ 
  Integration  over  the auxiliary field $f$ 
 gives
  \eqn\rpo{
 I=\int d\vp  \bigg[ \bet \dot \et  + \z \dot \z 
 +  \four (\bet \T\eta)^2 +   
 i \bet [  - \z \psi^m \D_m \T     + 
 ( A_m  \dot x^m - \ha \psi^m \psi^n F_{mn} ) ] \eta  \bigg]
\ .   }
 Finally, integrating over $\z$ we find 
 $$
 I=\int d\vp  \bigg[ \bet \dot \et  
 + \four (\bet \T\eta)^2  
  -  \four (  \psi^m \bet \D_m \T\eta ) \del_\vp^{-1}
  ( \psi^n\bet  D_n \T\eta) $$
  \eqn\repoy{ + \ 
  i \bet (  
  A_m  \dot x^m - \ha \psi^m \psi^n F_{mn} ) \eta \bigg]
\ .   }
Here  (cf. \tytr)
\eqn\inve{
\del_\vp^{-1} (\vp_1,\vp_2) 
= {1\ov \pi}\sum^\infty_{r=1/2 } 
{1\ov r  } \sin r  \vp_{12}  \  .  }
Note that the 
 $\d$-function 
defined on antiperiodic functions is 

\noindent
$\d^{(-)} (\vp_1,\vp_2) 
= {1\ov \pi}\sum^\infty_{r=1/2 } 
 \cos r  \vp_{12}$ so that (ignoring regularization, cf. \an)
  $K \cdot K = - \d^{(-)}$
 and  $\del_\vp \cdot \del_\vp^{-1} =
 \d^{(-)}, \ \del_{\vp_1} (\vp_1,\vp_2) \equiv 
  \del_{\vp_1} \d^{(-)}  
 (\vp_1,\vp_2)= - {1\ov \pi}\sum^\infty_{r=1/2 } 
{ r  } \sin r  \vp_{12} $.   
One could  think of using 
the regularized expression 
 $\del_\vp^{-1} (\vp_1,\vp_2) 
= {1\ov \pi}\sum^\infty_{r=1/2 } 
{e^{-r\ep} \ov r  } \sin r  \vp_{12}$, but that 
 leads to  complicated expressions as 
  it should be accompaneed by a similar regularization 
   in the $f^2$
  term  to preserve 1-d supersymmetry.

The resulting derivative expansion of $Z$ is thus 
expressed in terms of $\T, F_{mn}$ and their covariant 
derivatives.

In the abelian $U(1)$ case 
the integral over $\eta,\bar\eta$ in \repo\  is trivial
  and  the  tachyonic part of \repoy\  becomes equivalent 
  to the terms  
originally   derived  in 
  \rf{\hkm,\kmm} (cf. \zz) 
\eqn\abba{
 I=\int d\vp  \bigg( 
 \four \T^2  (x)
  -  \four  [ \psi^m \del_m \T(x) ] \del_\vp^{-1}
   [\psi^n  \del_n \T (x) ]  +   
  i [A_m (x)   \dot x^m - \ha \psi^m \psi^n F_{mn}(x) ]   \bigg)
  \ .   }  
It is easy to check directly that this action is invariant under
1-d supersymmetry\  $\delta x^m = \psi^m \varepsilon, \
  \delta \psi^m = \del_\vp x^m \varepsilon.$ 
Note  that if $\T$ is a constant 
 non-abelian 
  matrix then the path ordering  (the integral over $\eta$
   in \repo) is not relevant, and integrating over $f$ one gets 
   the potential factor  $\tr\  e^{ - {\pi \ov 2} \T^2}$ \kmm.
In general, however, 
the   non-abelian generalization of \abba\
  consistent with 1-d supersymmetry 
   is obtained  by  using 
  \repoy\
 (and not by adding trace with ordinary path ordering to 
 \abba).

  This 1-d supersymmetric theory defined by  \tytr,\abba\
\eqn\theo{
Z[\T,A] = \int d^D x  \ e^{-W} \ , \  \ \ \ \ \ \ \  e^{-W} =\ 
< e^{-I[x+ \xi, \psi]} > \ , }
$$ \ \ \ \ 
 <...> =\int [d\xi] [d\psi]  e^{ - { 1 \ov 4 \pi \a'}\int  ( \xi G^{-1}
 \xi 
 + \psi K^{-1} \psi) } \ , $$
    has  only logarithmic
    UV divergences,
   i.e. all  power divergences cancel out \rf{\mrt,\an}.
   This is  true  in the general 
   non-abelian case,  and also in the presence of  
   supersymmetric higher-derivative 
   interactions, and  is implied, e.g., 
   by  the non-local form 
   of the 1-d supersymmetric 
   lowest-dimension interaction in \repoy,\abba.

 In particular, there is no inhomogeneous $F$-dependent 
 term \tem\ in the  analog of \ta.
 Indeed, the coefficient  in the analog
 of \coo\   (i.e. in $Z$ computed for $F_{mn}=\const$) 
 now has both bosonic and fermionic contributions  
 and is finite as a result  \mrt:
 \eqn\cooe{
b_0= \sum^\infty_{n=1}  e^{-2\ep n} - 
 \sum^\infty_{r=1/2}  e^{-2\ep r}   = - { \te{1 \ov 2 }}+ O(\ep) \ . }
This  cancellation of power divergences
makes the  NSR  string partition  
function $Z[A, \T, ...]$ 
much better defined than in the bosonic string case.

One consequence is that 
the tachyon field manifestly decouples from the 
massless vector sector. 
This follows  of course from  conservation 
of G-parity ($\psi^m\to -\psi^m, \ \theta \to - \theta$)
under which  the tachyonic  vertex is odd, while the vector vertex
is even. In the S-matrix language, there are no tachyonic 
poles in the massless NS vector amplitudes
(so that the theory has of course 
consistent superstring truncation).
 Equivalently, this is obvious from 
\abba\ where $\T$ appears only quadratically.

As a result,  the subtleties like the one discussed 
in section 2.2 do not appear in the  NSR  case, 
and the renormalized  partition function 
$Z[A_\R, \T_R=0]$  
gives directly  the vector field effective action 
consistent  with the string S-matrix  and $\b^A$-function 
\an\ 
(renormalization of  logarithmic
divergences 
corresponding  to subtraction of  massless 
poles in the  string amplitudes is still 
needed in order to define the effective
action). 
  For example, it  was  demonstrated  in \an\ that  
the leading  derivative correction 
 to the BI term  in the partition function $Z$ \theo,\abba\
 which has the structure 
 $F^2 (\del \del F)^2 $ is  exactly the same 
 as in the action reconstructed from the 4-point NSR string  
 vector  amplitude.

\subsec{Tachyon  action and correspondence with  $\b$-function }
Let us compute the leading  $\T$-dependent 
terms in the abelian partition function 
 \theo\  using  derivative expansion.
Expanding in powers of  the quantum fields  $\xi,\psi$ 
 one finds
$$
I=\four \int d\vp  \bigg[ 
 \T^2 (x+\xi) 
  -   [\psi^m \del_m \T(x+\xi) ] \del_\vp^{-1}[
   \psi^n  \del_n \T (x+\xi)]  \bigg] $$ 
   \eqn\oip{
   = \four \int d\vp  \bigg[ \T^2 +
    (\T \del_m\del_n \T  + \del_m \T \del_n \T) \xi^m \xi^n 
   -  \psi^m \del_m \T  \del_\vp^{-1}
   \psi^n  \del_n \T   + O(\xi^3, \psi^2\xi) \bigg] 
  \ .   }   
The leading  one-loop contribution to \abba\
is thus ($2\pi \a'=1$)
\eqn\tiio{
W= {\te { \pi \ov 2}}  ( \T^2  + s_1 \T \del^2 \T  +
s_2 \del_m \T \del_m \T )   + O(\del^4) \ , }
where
\eqn\sea{
s_1 = G(\vp,\vp) = { \te { 1 \ov \pi}} \sum_{n=1}^\infty 
 {e^{-\ep n} \ov
n} = -{\te{ 1 \ov \pi}} \ln \ep + O(\ep) \ , 
 } 
\eqn\tio{
s_2 = (G +  K \cdot \del^{-1}_\vp)(\vp,\vp)=
{ \te { 1 \ov \pi}} ( \sum_{n=1}^\infty  {e^{-\ep n} \ov
n} - \sum_{r=1/2}^\infty  {e^{-\ep r} \ov
r} ) = -{\te { 1 \ov \pi}}\ln 4  + O(\ep) \ . }
[
The exact expressions for the sums are \an:

\noindent
 $\sum_{n=1}^\infty  {e^{-\ep n} \ov
n}=    - \ln ( 1 - e^{-\ep}) \ , 
 \ \  
\sum_{r=1/2}^\infty  {e^{-\ep r} \ov
r} = -\ln ( {  1 - e^{-\ep/2} \ov 1 + e^{-\ep/2} } ) $.]

Thus while the coefficient of $\del \T\del \T$ 
term is finite \rf{\kmm} \ 
(it  may probably  depend on a regularization only  if 
the latter  breaks 1-d 
supersymmetry, cf. \and),
the  coefficient of $\T \del^2 \T$
term is logarithmically divergent.
This divergence is to be renormalized  by  absorbing it into 
 $\T$.
Just as in the bosonic case, this 
logarithmic divergence determines the derivative term in the 
tachyon 
$\b$-function. The coefficient of the logarithmic pole
(i.e. the anomalous dimension of the tachyon vertex operator)
is of course the same as in the bosonic case, but the 
dimension of the  NS tachyon vertex
is  half of the bosonic one 
($\T \sim { 1 \ov \sqrt \ep}$). Thus   (cf. \bbb) 
\eqn\betta{
\beta^\T = - \ha \T - {\te{ 1 \ov 2 \pi} } \del^2 \T  \ . }
Introducing a constant $F_{mn}$ background 
means replacing $G$ by $G(F)$ in \exa\ 
and $K$ in \tytr\ by $K(F)$  \rf{\berg,\an}
\eqn\exta{
  K^{mn} (\vp_1,\vp_2|F) 
 =\  { 1 \ov \pi} \sum^\infty_{r=1/2} {e^{-\ep r}}
  \big[\G^{mn}(F) 
\sin r \vp_{12} + i \H^{mn} (F)\cos r \vp_{12} \big] \ .
 }
Then  one finds again \tiio\  but with 
the flat target space metric  replaced by $\G^{mn}(F) $
in \defi.
In particular, then \betta\ takes the form (see also \and)
\eqn\etta{
\beta^\T = - \ha \T - {\te{ 1 \ov 2 \pi} } 
\G^{mn} (F) \del_m \del_n \T  \ . }
Ignoring first the 
$F_{mn}$ background
and expanding $e^{-W}$  in derivatives  of the tachyon field
we obtain  the renormalized value 
of the partition function in \theo.
To simplify the expression 
let us  define the renormalized value of the tachyon $T$ by 
rescaling $\T$ by $\sqrt { \pi\ov 2}$. Then
\eqn\what{
Z[T] = c_0 \int d^D x \ e^{-T^2} \ 
\bigg[ 1 - \bar s_1  T \del^2 T - s_2 (\del T)^2 
+ O(\del^4 T) \bigg]
\ ,  }
where $\bar s_1$ stands for  a renormalized value of $s_1$ in \sea.
The same  expression, but without the $\bar s_1$ 
term was found in \kmm\ 
 where $T$ was taken to be linear in $x^m$ 
(i.e. $ \del^2 T$ was equal to zero).

It  is natural to expect  that in contrast 
to the bosonic string case  where one
 needs 
to shift $Z$ by a  derivative term \rf{\wi,\sa} 
  to  get the  action \rer\
reproducing the  tachyon beta function, in 
the NSR string case 
the (renormalized) 
partition function  is itself  
the correct action, not only in the massless vector
sector \an\ but also in the
tachyonic one \kmm.
To demonstrate that  $S[T]=Z [T]$ \what\ 
does indeed reproduce both terms in the 
 perturbative $\b$-function \betta\
it is crucial to include the $\bar s_1$ term in \what.
The   two  derivative-dependent 
  terms  in \what\  are closely related
through integration by parts (cf. \lea)
\eqn\wht{
S[T] = Z[T]= c_0 \int d^D x \ e^{-T^2} \ 
\bigg[ 1  +  c_1 \a' (\del T)^2 
  + c_2 \a' T^2  (\del T)^2  + O(\del^4 T) \bigg]
\ ,  }
 \eqn\coff{
 c_1 = 2\pi (\bar s_1- s_2 ) \ , \ \ \ \ \ \ \ \  
 c_2 = - 4\pi  \bar s_1\ . } 
Since $s_1$  was logarithmically divergent
before renormalization, its renormalized 
 value $\bar s_1$ is, in principle, 
ambiguous and, as in the bosonic case
(see \lea--\rwt) can be tuned 
to match  the variation of $S[T]$ with 
the $\beta^T$-function in \betta.
Indeed, we find
$$
 {\delta S \ov \delta T}
 =2 \ c_0\ e^{ -  T^2} \ 
 \bigg[-  T  - c_1 \a' \del^2 T 
$$  \eqn\rwty{ - \ c_2 \a'T^2\del^2 T
  + (c_1 - c_2) \a'  T (\del T)^2  
   +   c_2 \a'  T^3 (\del T)^2  + O(\a'^2\del^4 T) 
  \bigg] \ . }
 Thus the linear terms  here are proportional to 
  \betta\ if  $\bar s_1= { 1 \ov \pi} ( 1-\ln 4)$, i.e. if  
  \eqn\seaa{
  c_1= 2 \ . }
  Then  $c_2= 4 (\ln 4-1) > 0 $ so that the kinetic term in \wht\
  is positive for all $T$.

While  it may seem making  little sense to try 
to reproduce the  correct tachyonic  mass  plus  kinetic  terms
in the action 
using the  perturbative derivative expansion, 
the point is that  the freedom of field redefinitions
allows one to do that, both in the bosonic
\rf{\sha,\gersha}
 and in the NSR 
cases. The resulting field space ``metric" 
 $\kappa$  is then simplest in such scheme.
 

The generalization of the action \wht\ to the presence 
of a $F_{mn}=\const$   background is 
straightforward (cf. \leaks)
(the partition function for
$\del_m T=\const, \ F_{mn}=\const$ background 
was computed in \and; its expansion in $\del T$ reproduces
part of the $c_1$ 
term in the expression below
which  corresponds to the $s_2$ term in \what)
$$
S= c_0\int d^D x \ e^{ -  T^2}  \sqrt 
{\det (\d_{mn} + 2\pi \a'  F_{mn})} 
 \ \bigg[\ 1 +\  c_1  \a' \G^{mn} (2\pi \a' F) 
 \   \del_m T \del_n T $$ \eqn\eaks{
  + \     c_2 \a' \G^{mn} (2\pi \a'F)\   T^2  \del_m T \del_n T 
 \  +\       O(\a'^2 \del^4 T,\a'^2 \del^2F) \bigg] 
 \ ,   }
where we restored the  dependence on $\a'$.
This action is consistent with \etta\ for 
$c_1 =2$.

The non-abelian generalization of \eaks\
may be obtained, in principle,  from the gauge-invariant 
path integral defined by   \repoy.


\newsec{Closed string  theories}

In bosonic closed string  case we start with the \sm \ft\ (for simplicity, 
we shall ignore the Kalb-Ramond 
antisymmetric tensor coupling which is not essential
for the present discussion and can be easily included)
\eqn\siga{
I =
 \int d^2 z \sqrt g  \bigg[ \ep^{-2} T_0(x) + {\te 
 { 1 \ov 4 \pi \a'}} \del^\m x^m \del_\m x^n G_{mn} (x) 
 + {\te { 1 \ov 4 \pi }} R^{(2)} \pp(x) \bigg] \ . 
 } 
This  model is renormalizable within $\a'$ expansion.
The corresponding bare
partition function on 2-sphere has the form \rf{\ft}
\eqn\crita{
Z[T_0(\ep),G(\ep),\pp(\ep),\ep]
 = d_0 \int d^D x\  \sqrt G\  e^{-2\pp -  \ep^{-2} A T_0 }\  e^{-W} \ , }
where we shifted $x(z) \to x + \xi(z)$  so that (as in \qre) 
$W$ is given by the path integral over the non-constant 
fluctuations $\xi$ (see \rf{\ft,\TT,\mep} for details). The coefficient
$A$ is  the area of  $S^2$  with   fiducial 2-d metric (it can 
be absorbed into the renormalized value  of $T_0$). 
 The leading  logarithmically divergent terms in $W$ are 
 found to be \rf{\poll,\ft} 
$$ W =   \ha \gamma \ln \ep  +  O(\ln^2\ep) + {\rm finite} \ , \ \ \  $$ 
 \eqn\loga{ \g = \cc_0 -  \a' \cD^2 ( \ep^{-2} A T_0)
  -  2 \a' \cD^2 \pp   - \a' R  + O(\a'^2) \ , \ \ \ \ \ \ \ \ 
  \cc_0 =   {\te{ 2 \ov 3} }(D- 26)\ . } 
  For example, taking the derivative of $Z$ over 
  $\ep$ reproduces the standard
  perturbative  closed string tachyon
  $\b$-function, 
  \eqn\tacha{
  \beta^T = - 2 T  - \ha \a' \cD^2 T \ ,  } 
  with the corresponding  Weyl anomaly coefficient being
  $\bar \beta^T =\beta^T + \a' \del^m \pp \del_m T$.
  
To obtain the effective action for the  massless 
fields $S[G,\pp]$ from the partition function $Z$
 one should, as in the  open string case, renormalize the logarithmic 
 infinities which corresponds to subtracting massless 
 poles in the string amplitudes \TT.  An additional subtlety
 of the closed string case is that the M\"obius group volume 
 has logarithmic divergence, and it should be subtracted, in the RG invariant
 way,  
 by applying $\del \ov \del \ln\ep$ to the bare  value of $Z$ 
 \mec.
To
 compare  the formal generating functional 
 $Z$ to  massless  effective action  reconstructed from string amplitudes
 one needs
to subtract both 
 M\"obius infinities and massless poles (UV logarithms). 
While the former are power-like in the open string case,
they are also logarithmic in the closed string case.
That means that ``extra" log should be subtracted from Z
in the closed  string case.

 Expressed in terms of renormalized couplings, 
 the effective action is then \mec\  ($\l^i = (T, G, \pp)$) 
 \eqn\eeqq{
 S = - ({\del Z \ov \del \ln \ep})_{\ep=1} 
 =  \beta^i\cdot { \delta Z \ov \delta \l^i}  \ .  }
 Because of the diffeomorphism invariance, 
 the  expression for 
 $S$ can be written  also as $ S =
 \bar  \beta^i\cdot { \delta Z \ov \delta \l^i}$, 
 where $\bar \b^i = \b^i + (\delta \l^i)_{\a' \del \pp}  $ 
 (i.e. 
 $\bar \b^\pp = \b^\pp + \a' \del^m  \pp\del_m \pp, \ \ 
   \bar \b^G_{mn}  = \b^G_{mn}  + 2\a' \cD_m\cD_n \pp $, etc.) 
         are the Weyl anomaly coefficients, the vanishing of which 
  should be  equivalent to the conditions of 
  stationarity of the action.
   
 Explicitly, 
 $$  
 S=    \int d^D x\ \sqrt G \ e^{-2\pp - T }\ 
  (- 2 T + \ha  \g )   $$   \eqn\teew{ =\ 
  \int d^D x\ \sqrt G  e^{-2\pp - T } \bigg[
  \ha \cc_0 - 2 T -  \ha  \a' \cD^2 T 
  -   \a' \cD^2 \pp   - \ha  \a' R  + O(\a'^2) \bigg] \ ,  }
 where $T$ is a  renormalized value 
 of the tachyon rescaled by $A$.
 Equivalently, 
   $$  S = 
  \int d^D x \sqrt G  e^{-2\pp - T } \big (\b^T + 
  2 \b^\pp - \ha G^{mn}  \b^G_{mn} \big) $$ 
 \eqn\nuew{  = \ \int d^D x \sqrt G  e^{-2\pp - T } \big (\bar \b^T + 
  2\bar \b^\pp - \ha G^{mn} \bar \b^G_{mn} \big) , 
  }
  i.e. 
  \eqn\voot{
  S= 
   - \big(  { d  \ov  d  \ln \ep } \Z  
  \big)  _{\ep=1} \ , \ \ \ \ \ \ \   \ \ \ \ 
   \Z  = \int d^D x\ \sqrt G  \  e^{-2\pp - T }  \ . }
Note that (in contrast to the open string case \subs,\subse)
 the coefficients of derivative terms here are scheme-independent.
 Setting the tachyon to zero  (and integrating by parts) 
  we get the standard  closed string 
 effective   action consistent with S-matrix \nss\
 and massless $\beta$-functions 
 \rf{\hon,\call} 
 \eqn\teyew{
 S=-  \ha   \int d^D x\ \sqrt G  \  e^{-2\pp }
  \bigg[
  -\cc_0  
  +   4 \a' (\del \pp)^2    +  \a' R  + O(\a'^2) \bigg] \ .  }
  As discussed  in \mec, the definition \eeqq\ in general
  leads  to $ S $  given by  the  space-time 
integral of the 
``central charge" coefficient.

 The  functional \teew\ is not, however, 
 the right action for $T\not=0$:   it
 does not have  the standard perturbative vacuum 
 ($ D=26,T=0, \pp=\const, G_{mn}=\d_{mn}$) as its stationary point
 (equivalently, the tachyon tadpole on 2-sphere does not vanish
 even after taking the derivative in \eeqq).
 Another indication of a problem  in  \teew\ is that 
 one can almost completely absorb $T$ into the dilaton:
 introducing  $\td \pp = \pp + \ha T$ 
 one is left with  only  a linear term
 in $T$. Futhermore, the kinetic term of $T$ in
 \teew\ ($- \ha \a' \del^m T \del_m  T$ after integration by parts)
 has apparently  the ``wrong" sign.
We use the 
 Euclidean signature so the  correct sign
 for a scalar kinetic term is plus. This is the sign 
 of the dilaton kinetic term  in  \teew\ 
 after redefining the metric to decouple graviton from dilaton
 (i.e. after going to the Einstein frame).
 In  general, $T$ and $\pp$ and the graviton
 are  mixed,  so that  their kinetic matrix is to be
 diagonalized  before discussing the signs  (see below).

 As in the open string case, to try to  find
   a consistent  action
 that reproduces  $\bar \beta^i=0$ conditions for all the three
 fields 
  one is  thus to subtract the tachyon tadpole in an RG invariant way.
 By analogy with  \rer,  let us  define
 \eqn\qooeq{
 \hat S= S +\ha  \beta^i \cdot  {\delta S  \ov \delta \l^i } =
  \beta^i \cdot  {\delta  \ov \delta \l^i } Z   
  + \ha   \b^j \cdot {\delta  \ov \delta \l^j } 
   ( \b^i \cdot {\delta  \ov \delta \l^i }  Z) \ . 
  }
The  coefficient 1/2
 accounts for the factor of two difference in dimensions
 of the open and closed  strings tachyons, cf.  \bbb,\tacha.
 Note that since $S$ in \eeqq\ is RG invariant
 ($ {d \ov d\ln \ep } Z=0 \to {d \ov d\ln \ep}
  {\del \ov \del \ln \ep}
  Z=0 $), the same applies to
 each of the two  terms in $\hat S$.

 Expanding near  the standard 
  flat string vacuum ($D=26, \ G_{mn}=\const, 
  \  \pp=\const$)  one is to 
  keep the dilaton and graviton  perturbations
in $\hat S$  and  consistently 
decouple them from $T$  by field redefinitions
(we 
are grateful to
S. Frolov for an important discussion of   this point).
Observing
 that  for small perturbations near the flat  vacuum 
$\beta^\pp = - \ha\a' \del^2 \pp, \  \beta^G_{mn} =\a' R_{mn} , $
one finds 
$$ \hat S= S + 
\ha (  \beta^T \cdot { \delta  \ov  \delta T } 
+ \beta^\pp \cdot { \delta  \ov  \delta \pp }
+ \beta^G_{mn} \cdot { \delta  \ov  \delta  G_{mn} }) S  $$
 \eqn\teewa{ =\ 
  \int d^D x\ \sqrt G  e^{-2\pp - T } 
  \bigg[  - 2 T^2  + \ha\a'  \cD^2 T - \a'  T \cD^2 T  
  -  \ha  \a' (1 + 2 T) (R + 2 \cD^2 \pp )    + O(\a'^2) \bigg] \ .   }
This action no longer has a  tadpole $T$-term (cf. \teew),  
but to decouple  graviton  from  scalars we
still
 need, as  usual, 
to redefine the metric. Ignoring ``mixed" (tachyon--massless)
terms which are of 
higher than quadratic order in the fields we can approximately 
replace 
$e^{-2\pp - T }(1 + 2 T)$ factor in front of $R$  by $e^{-2\vp }$, 
\ 
$\vp \equiv \pp - \ha T$  and  set 
$G_{mn} = e^{ 4\vp\ov { D-2} } g_{mn}$.
Then 
$ \int d^D x \sqrt G  e^{-2\pp - T } 
(1 + 2 T) (R +2 \cD^2 \pp ) \  \to \  
\int d^D x \sqrt g  \big[ R(g)
 - { 4\ov { D-2} }( \del \vp)^2 + 2 \del^m T \del_m \vp 
\big]$.  As a result,  the action \teewa\ takes the form
 \eqn\foru{ \hat S 
=  \int d^D x\ \sqrt g \ \bigg(  e^{- 2T }
  \big[  - 2 T^2 e^{ 4 \vp\ov D-2} 
   + 2 \a' (1-T) ( \del T)^2  
 \big]  - \  \ha  \a' \big[R(g) - {\te{ 4\ov { D-2} } }( \del \vp)^2 
  \big]    + O(\a'^2) \bigg) \ .    }
If one does the exact 
redefinition  
 $ (1 + 2 T)e^{-2\pp - T } = e^{-2 \p}$, i.e.
 $\p= \pp + \ha T - \ha \ln (1 + 2T)$, 
 and thus   $G_{mn} = e^{ 4\p\ov { D-2} } g_{mn}$, 
 then 
 one finds the following action
 $$  \hat S 
= \int d^D x\ \sqrt g \ \bigg(  
    - { 2 T^2\ov 1 + 2 T}  e^{ 4 \p\ov D-2} 
   + { 2\a' \ov (1 + 2 T)^2} ( \del T)^2 
 -  \ha  \a' \big[R(g) - {\te{ 4\ov { D-2} } }( \del \p)^2
 \big]    + O(\a'^2) \bigg) \ .   $$
 which is equivalent to the one above  in the 
 quadratic order in $T$.
 Introducing $\td T = \ha \ln (1 + 2 T)  $ 
 (assuming $T > - \ha$) 
  its tachyonic part becomes simply 
 $  \hat S 
= \int d^D x\ \big[
    - 2 \sinh^2 \td T  + 2 \a' (\del \td T)^2 \big] .
    $

Ignoring the graviton and dilaton 
terms (decoupled  to quadratic order in fluctuations) 
the tachyon part of the action is thus
\eqn\foqu{\hat S 
= \ \int d^D x \ e^{- 2T }\ 
  \big[  - 2 T^2  
   + 2 \a' (1 -T)  (\del T )^2   + O(\a'^2) \big] \ .    }
This action has a  structure similar to \lea, 
but  the  value of the coefficient 
of the second kinetic  term, though now  positive, 
 is not the one   needed 
 to reproduce the   $\b^T$-function \tacha.
It may be that the definition \qooeq\ still needs some 
further refinement.

The  potential term in \foqu\ $V= - 2 e^{-2T}  T^2$
has tachyonic maximum  at $T=0$ and the stable minimum
at $T= 1$. However, that minimum is not reached  as the kinetic
term  of $T$  changes sign at $T=1$.
In general, in discussing the vacuum structure
one should take into account 
 a non-trivial mixing of the tachyon 
with the 
dilaton and the metric. For example, 
for linear dilaton and $D=2$ the tachyon should be massless
(as follows from $\bar \beta^T=0$), and  
 in this case one should  expect to 
 find no potential term.

The exponential potential in \foqu\
 may be 
 suggesting (by analogy with the open string case)
   a possibility of $T$ rolling to infinity along some
 directions, and  such behavior  
  should be accompaneed by a non-constant  dilaton 
  to preserve the central charge condition 
  (see  also \hkm\ for related  remarks).

In the closed  NSR string case the  tachyon
vertex  has the  following form (in the 0-ghost picture) 
  $ \int d^2 z \  \psi^m \bar \psi^n \del_m\del_n T$.
Its   2-d supersymmetric
generalization is 
$ \int d^2 z d^2  \theta\  T( \h x) $,  where 
$\h x^m   = x^m  + \theta \psi^m  + \bar \theta \bar \psi^m  + 
\bar\theta  \theta  f^m$.
Combined with the  kinetic   term  
 $ \int d^2 z d^2 \theta\   D \h x^m \bar D \h x^m $
it leads, after the elimination of the auxiliary field $f^m$, 
 to the following tachyon terms in the \sm action 
 (which are the familiar  superpotential  terms in $N=1$ 
 supersymmetric
 scalar 2-d field theory, cf. also \abba)
 \eqn\typel{
 \int d^2 z \big[ \del^m T(x)  \del_m T(x)
  +  \psi^m \bar \psi^n  \del_m \del_n  T(x) \big] \ . }
As in the open  NSR string case, the resulting partition function
and  thus the effective action \eeqq\ it generates
  are   even in $T$.
   As a result, there is no need for an additional subtraction
like \qooeq. Since the \sm depends on $T$ only through its derivatives, 
there is no tachyon potential term in $S$.

The leading $T$-dependent logarithmic divergence in  $Z$ 
comes from  the expansion of  $\del_m T \del^m T$
term in \typel\ 
and corresponds to the $\b$-function 
$( - 1  - \ha \a' \cD^2 )  T$. 
 The leading $G_{mn}$ and $\pp$-dependent terms in
  $Z$ and $S$  in the NSR case 
are the same as in the bosonic case 
 \ame\ (with the obvious
  replacement of $\cc_0$ in \loga\ by $D-10$). 
While the NS-NS part of the   
effective action generated by the \sm 
appears to depend on $T$ only
through its derivatives, functions of $T$ may still
be present in the R-R sector (where one is to use the 
ghost -1 tachyon vertex \kle).

\bigskip
\noindent {\bf Acknowledgements}

We are grateful to   
O. Andreev, S. Frolov,  D.  Kutasov and 
S. Shatashvili for 
 useful  discussions and remarks. 
This work was  supported in part by
the DOE grant  DE-FG02-91ER40690, 
INTAS grant No. 99-0590
and PPARC SPG grant  PPA/G/S/1998/00613.

\listrefs
\end